\documentclass{aastex} 

\RequirePackage{color}

\newcommand{\emaila}{jmould@swin.edu.au}

\begin{document}
\title{BLACK HOLES IN 4 NEARBY RADIO GALAXIES}
\shorttitle{Black holes}
\shortauthors{Mould et al.}


\author{Jeremy Mould$^{1}$, Tony Readhead$^{3}$, Garret Cotter$^4$, David Batt$^{2}$ and Mark Durr\'{e}$^1$}


\altaffiltext{1}{ Centre for Astrophysics and Supercomputing, Swinburne University of Technology, Melbourne, Victoria 3122, Australia}
\altaffiltext{2}{ School of Physics, University of Melbourne, Parkville, Victoria 3010, Australia}
\altaffiltext{3}{California Institute of Technology, CA91125, USA}
\altaffiltext{4}{Department of Physics, University of Oxford, Denys, UK}
\email{\emaila}

\begin{abstract}
We study the velocity dispersion profiles of the nuclei of NGC 1326, 2685, 5273 and 5838 in the CO first overtone band. 
There is evidence for a black hole (BH) in NGC 1326 and 5838. Gas is seen flowing out of the nuclear region of NGC 5273.
We put upper limits on the nuclear BHs responsible for its activity and that of NGC 2685.

\end{abstract}

\section{Introduction}
Understanding activity in galactic nuclei requires high spatial resolution.
Kormendy \& Richstone (1995) have outlined the techniques for quantifying the supermassive
black holes that power active galactic nuclei (AGN).
Our strategy (Mould et al 2012) is good seeing infrared spectroscopy of AGN
in a volume limited sample, followed by adaptive optics spectroscopy on large aperture telescopes.
In this paper we present Palomar TripleSpec spectra of a number of nearby radiogalaxies
of early type.

NGC 1326 is a ring barred S0 galaxy in the Fornax cluster with circumnuclear star formation (Buta et al 2000).
Our second galaxy is a Hubble Atlas polar ring galaxy, an S0 Seyfert 2. Schinnerer \& Scoville (2002) detected four giant molecular cloud
associations within the polar ring in NGC 2685 (the Helix) with of order 10$^7$ M$_\odot$ of molecular hydrogen.
Dust has been detected with Spitzer in our third S0 galaxy, NGC 5273, totalling 2.5 $\times$ 10$^5$ M$_\odot$
by Martini et al (2013). NGC5838 has a nuclear star cluster of 5 $\times$ 10$^7$ M$_\odot$ (Scott \& Graham 2013).

\section{Sample and observations}
We have drawn our radiogalaxy sample from Brown et al (2011), further limiting the distance to 20 Mpc
in order to have 100 pc resolution in 1$^{\prime\prime}$ seeing.
Observations of NGC 1326, 2685, 5273 \& 5838 were obtained on the Hale Telescope in 2011 and 2012. 
Obtaining our Palomar TripleSpec spectra was described by Mould et al (2012) and data reduction
was outlined by Batt et al (2014, Paper I). We very briefly recap this here. The spectrograph
has resolution of 2600 with a 1$^{\prime\prime}$ slit, and observations were made with the nucleus in 
two slit positions ABBA in 4 $\times$ 5 minutes. These were followed by observations of an A0
star for telluric correction and seeing measurement. Flatfielded spectra were subtracted and extracted at different impact
parameters along the slit, yielding the wavelength shifts and first overtone CO line widths
given in Table 1. The IRAF cross-correlation task $fxcor$ was used for this purpose with the Gemini
library stellar template HD2490 interpolated to the same resolution.

Table 1 gives the radial position of the extracted spectrum in column (1), the pixel shift between that and the template
 in column (2), the peak height of the cross-correlation in column (3) and the FWHM of the fit to the cross-correlation
in column (4). The units of columns (2--4) are pixels.

\begin{deluxetable}{llllllll}
\tabletypesize{\small}
\tablecaption{Raw crosscorrelation data}
\tablehead{\colhead{}&
\colhead{}&
\colhead{}&
\colhead{}&
\colhead{}&
\colhead{}&
\colhead{}&
\colhead{}}

\startdata

{\bf NGC 2685}\\
	position&	pixel shift&	peak&	fwhm&
	position&	pixel shift&	peak&	fwhm\\
   (arcsec)&[2]&[3]&[4]&(arcsec)&[2]&[3]&[4]\\
	0	&	-189.96	&	0.23&	15.4&
	0	&	-190.55	&	0.21&	12.16\\
	0	&	-190.68	&	0.26&	14.96&
	0	&	-190.7	&	0.25&	15.1\\
	0.9	&	-190.5	&	0.26&	14.2&
	1.63	&	-187.09	&	0.24&	14.8\\
	0.73	&	-188.4	&	0.27&	13.9&
	0.79	&	-188.99	&	0.18&	14.98\\
	0.79	&	-190.65	&	0.26&	13.41&
	0.84	&	-189.96	&	0.2&	12.7\\
	0.79	&	-191.01	&	0.2&	12.3\\  \\
{\bf NGC 5838}\\
	position&	peak height&	fwhm&	shift&
	position&	peak height&	fwhm&	shift\\
   (arcsec)&[2]&[3]&[4]&(arcsec)&[2]&[3]&[4]\\
	0	&0.366	&27.4&	-0.76& 
	0	&0.519	&29.1&	-0.67\\	
	0.316	&0.329	&27.1&	-1.31&
	0.632	&0.301	&13.1&	-0.1\\
	0.948	&0.333	&14.2&	-0.6&
	1.264	&0.368	&10.2&	-0.41\\
	1.58	&0.331	&10&	-0.27&
	1.896	&0.314	&8.18&	-0.08\\
	2.212	&0.322	&11.4&	-0.28&
	0.316	&0.325	&13.1&	0.55\\
	0.632	&0.391	&8.35&	-0.03&
	0.948	&0.398	&8.99&	0.249\\
	1.264	&0.476	&10.6&	0.01&
	1.58	&0.462	&12.3&	0.4\\
	1.896	&0.484	&18.6&	0.89&
	2.212	&0.503	&14.3&	1.01\\	
	0.316	&0.397	&13.7&	-0.13&
	0.632	&0.398	&14.6&	-0.4\\
	0.948	&0.394	&11.7&	-0.71&
	1.264	&0.39	&13.9&	-0.49\\
	1.58	&0.418	&8.8&	-0.47&
	1.896	&0.381	&10.1&	-0.52\\ \\
{\bf NGC 5273}\\
	position&	peak&	fwhm&	position&	peak&	fwhm\\
   (arcsec)&[2]&[3]&(arcsec)&[2]&[3]&\\
	0	&	0.28&	28.36& 
	0	&	0.22&	15.6\\
	0	&	0.28&	23.22&
	0	&	0.22&	20.63\\
	1.57	&	0.13&	21.09& 
	0.73	&	0.26&	17.12\\
	1.99	&	0.23&	16.76&
	1.09	&	0.23&	24.18\\
	1.58	&	0.12&	11.94&
	0.79	&	0.28&	20.57\\
	1.69	&	0.12&	14.22&
	0.9	&	0.2&	11.31\\
	1.18	&	0.18&	21.12& 
	1.97	&	0.16&	7.22\\
	0.54	&	0.19&	22.96&
	1.33	&	0.1&	21.68\\
	0.84	&	0.29&	27.98&
	1.63	&	0.13&	15.2\\
	1.07	&	0.24&	24.97&
	1.86	&	0.19&	14.78\\   \\
{\bf NGC 1326}\\
	position&	peak&	fwhm&	position&	peak&	fwhm\\
0&0.4&20.61&
0&0.39&20.56\\
0&0.4&20.13&
0&0.42&20.84\\
0.78&0.43&22.4&
0.56&0.41&21.94\\
1.62&0.43&22.17&
1.12&0.38&20.38\\
1.01&0.42&22.8&
0.95&0.43&21.99\\
1.57&0.41&22.61&
1.8&0.39&24.8\\
0.45&0.42&21.96&
0.62&0.44&21.96\\
1.41&0.41&20.65&
1.07&0.42&21.61\\
0.9&0.41&20.88&
0.68&0.42&21.4\\
1.52&0.39&20.5&
1.52&0.41&22.17\\ 
\enddata
\end{deluxetable}

\section{Kinematics and dynamics}
\subsection{NGC 1326}
NGC 1326 has been imaged by the Hubble Space Telescope (Figure 1) and
its ultraviolet light distribution 
is displayed in the radial profile from the IRAF STSDAS surface photometry task
$ellipse$ in Figure 2.

The Jeans equation allows us to predict the velocity dispersion profile $\sigma$(r) corresponding to this light 
distribution, assuming spherically distributed stars on isotropic orbits. To do this, we need the logarithmic derivatives
 with respect to radius of the density and velocity dispersion profiles. The former is obtained numerically using an 
Abell transform, the latter by calculating the (small) slope of the velocity dispersion data. The visual mass to light 
ratio is a free parameter in this model and we fit it to the data at r $>$ 80 pc, finding M/L = 6.5 in solar units, 
a normal value for a stellar population not dominated by dark matter. TripleSpec line width values were normalized 
in the same way as in Paper I.

The addition of a 1 $\times$  10$^7 M_\odot$ 
black hole modifies the mass distribution and $\sigma$(r). 
It is a better fit to the data than the solid line in the lower part of Figure 2. The no BH model is ruled out with 70\% confidence based on $\chi^2$.

\subsection{NGC 2685}
NGC 2685 is a polar ring galaxy, known as `the spindle'. The HST nuclear image
is reproduced in Figure 3 and the light distribution has been fitted
with a `nuker profile' (Lauer et al 2007). The profile appears in Figure~4,
the model fit, and
$\chi^2$ per degree of freedom implies that M$_\bullet~>$ 3 $\times$ 10$^7$ with less
than 20\% probability.
This is consistent  with Beifiori et al (2009), who find an upper limit 
M$_\bullet~<$ 1.1 $\times$ 10$^7M_\odot$. The innermost datapoint
has been located, not at zero radius as Table 1 would imply, but at the effective light
centre of the zero radius observation taking account of seeing.

\subsection{NGC 5273}
We fitted a nuker profile to archival HST WFPC2 PC data (Figure 5), obtaining ($\alpha, \beta, \gamma$)
= (1.8, 1.8, 0.75) and normalized the profile to the surface photometry of Mu\~noz Marin et al (2007) with r$_b$ = 50 pc. Figure 6 is the model fit, and
$\chi^2$ per degree of freedom implies that M$_\bullet~ >$ 10$^8$ with less
than 25\% probability.
We assumed the Tonry et al (2001) surface brightness fluctuations  distance of m-M = 31.09 $\pm$ 0.26. 

The TripleSpec spectrum also shows an interesting He I 10830\AA~ line (Figure 10). Silhouetted against the broad line region helium emission and
its luminous (10$^8$ L$_\odot$) x-ray gas (Liu 2011) is a P-Cyg profile of cooler
(kT $\sim$ 30 eV) neutral gas with a terminal outflow velocity of 750 km/sec. This object will
repay IFU study of its circumnuclear gas and modelling to determine the outflow rate.

\subsection{NGC 5838}
Calculation of a predicted stellar velocity dispersion profile was described for galaxies with nuker
profiles in Paper I. NGC 5838 has such a profile (Lauer et al 2007).
Figure 7 shows the nucleus of NGC 5838 and Figure 8 shows a fit with M/L = 30
and a black hole of 1 $\times$ 10$^8$ M$_\odot$. Note that Lauer et al assume V-H = 2.39 in converting NICMOS data
to visual magnitudes. We also adopted their distance of 22.2 Mpc. The no BH model
is rejected with 98\% confidence.

\section{Summary}
We summarize our findings in Table 2. In two cases we have SMBH detections; in two cases
we have upper limits on the SMBH mass. Our upper limit for NGC 5273 is consistent with
the result from reverberation mapping of 4.7 $\pm$ 1.6 $\times$ 10$^6~M_\odot$ by Bentz et al 2014.
Figure 9 shows our 4 radio galaxies in their Magorrian diagram. NGC 5838 is
plotted at $\sigma$ = 290 km/sec (McElroy 1995).

\vspace*{1 cm}

\centerline{\bf Table 2: Black hole masses}
\begin{tabbing}
Namessss\=Typess\=Distances\=Msssssss\=ssssss\=ss\kill
Name\>Type\>Distance\>M$_V$\>M/L\>SMBH\\
NGC\>\>(Mpc)\>\>\>M$_\odot$\\

N1326\>SB0+\>20.5\>--21.05\>6.5\>1 $\times$~10$^7$ \\
N2685\>SB0+\>14.3\>--19.72\>1.3\>$<~3^*\times~10^7$\\
N5273\>S0\>20\>--20.1\>1$^\dagger$\>$<$ 1 $\times$ 10$^8$\\
N5838\>S0-\>22.2\>--20.51\>30\>1 $\times$ 10$^8$ \\
~ *1.1 (Beifiori et al 2009)\>\>\>\>$^\dagger$ UV M/L\\
\end{tabbing}

\acknowledgements
We thank our referee for comments that improved the paper.
We are grateful for the support of the Australian Research Council through DP140100435. GC acknowledges support from STFC grant ST/K005596/1.
Spectra were extracted using a version of the Spextool program modified for the Palomar TripleSpec Spectrograph
(Cushing et al. 2004, ; M. Cushing, private communication 2011). We acknowledge the Hubble Legacy Archive, a facility of STScI,
which is operated by AURA for the National Aeronautics and Space Administration (NASA).
This research has made use of the NASA/IPAC Extragalactic Database (NED) which is operated by the Jet Propulsion Laboratory, California Institute of Technology, under contract with NASA. This research has also made use of IRAF, software written by NOAO and data products from the Gemini Observatory, which are operated
by AURA under a cooperative agreement with NSF. David Batt was a summer student at Swinburne University while
this work was carried out.

\vfill\break
\begin{figure}[t]
\vspace*{-5 cm}
\includegraphics[scale=.25]{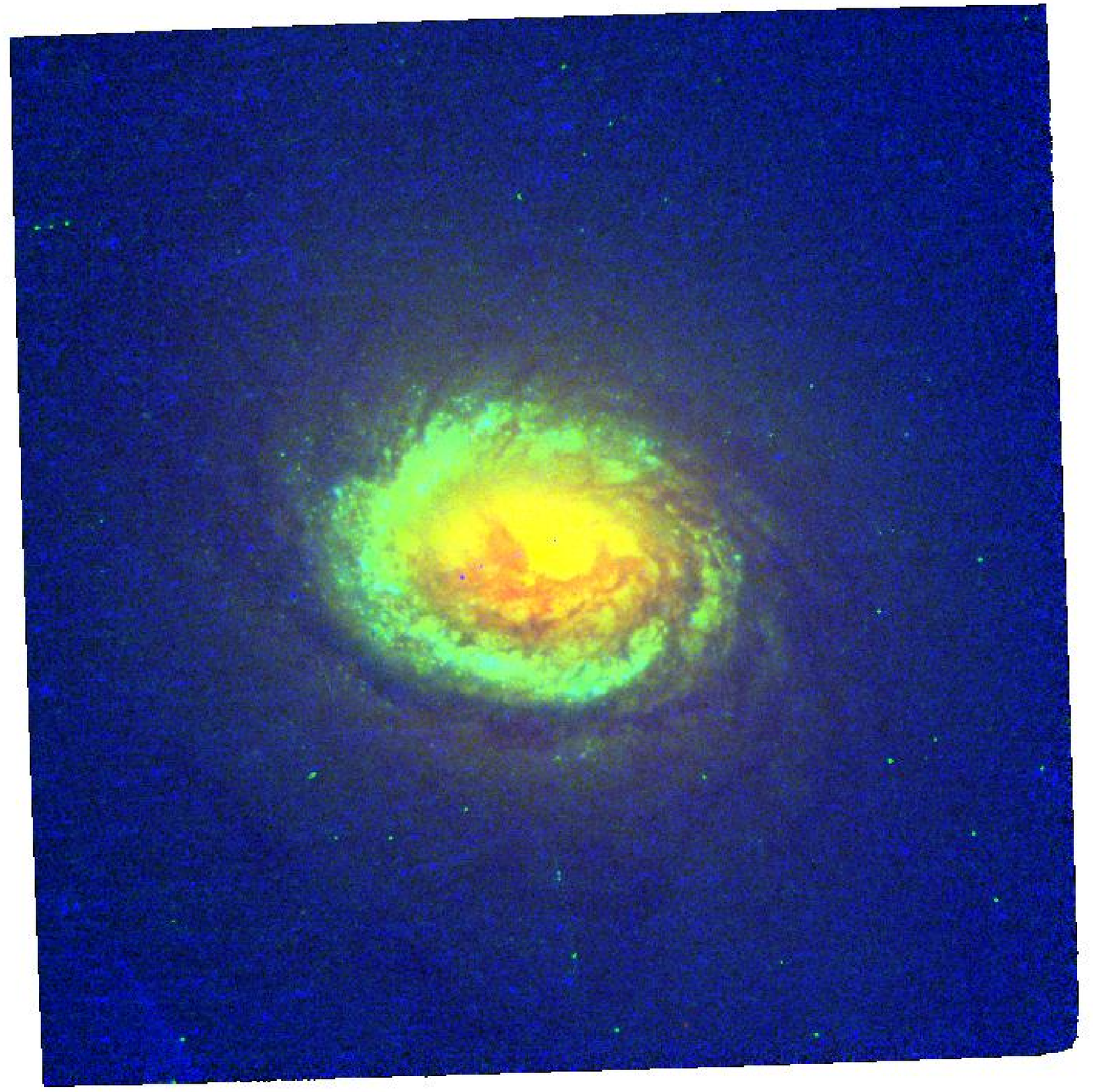}   
\caption{The HST WFPC2 PC image of the nucleus of NGC 1326, fov 36$^{\prime\prime}$, orientation N up and E to the left. 
The filters in RGB order are F814W/F439W/F255W.} 
\includegraphics[width=.7\columnwidth,angle=-90]{n1326.eps}   
\caption{$top$ UV radial surface brightness profile of NGC 1326 converted to V band by subtracting the NED U-V colour 1.15; $bottom$ E and W velocity dispersion profile
distinguished by the symbols. The solid line is a fit to the data with M/L = 6.5. The dashed line supposes the
presence of a 10$^7$ M$_\odot$ black hole. Velocity dispersion uncertainties are a similar siz to the plotting symbols.}
\end{figure}

\begin{figure}[t]
\vspace*{-3 cm}
\includegraphics[scale=.17]{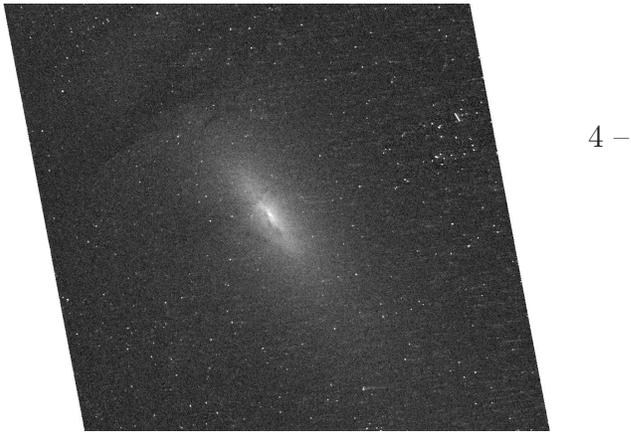}   
\caption{The HST ACS HRC image of the nucleus of NGC2685. The filter is F330W. }
\end{figure}
\begin{figure}
\includegraphics[width=.7\columnwidth,angle=-90]{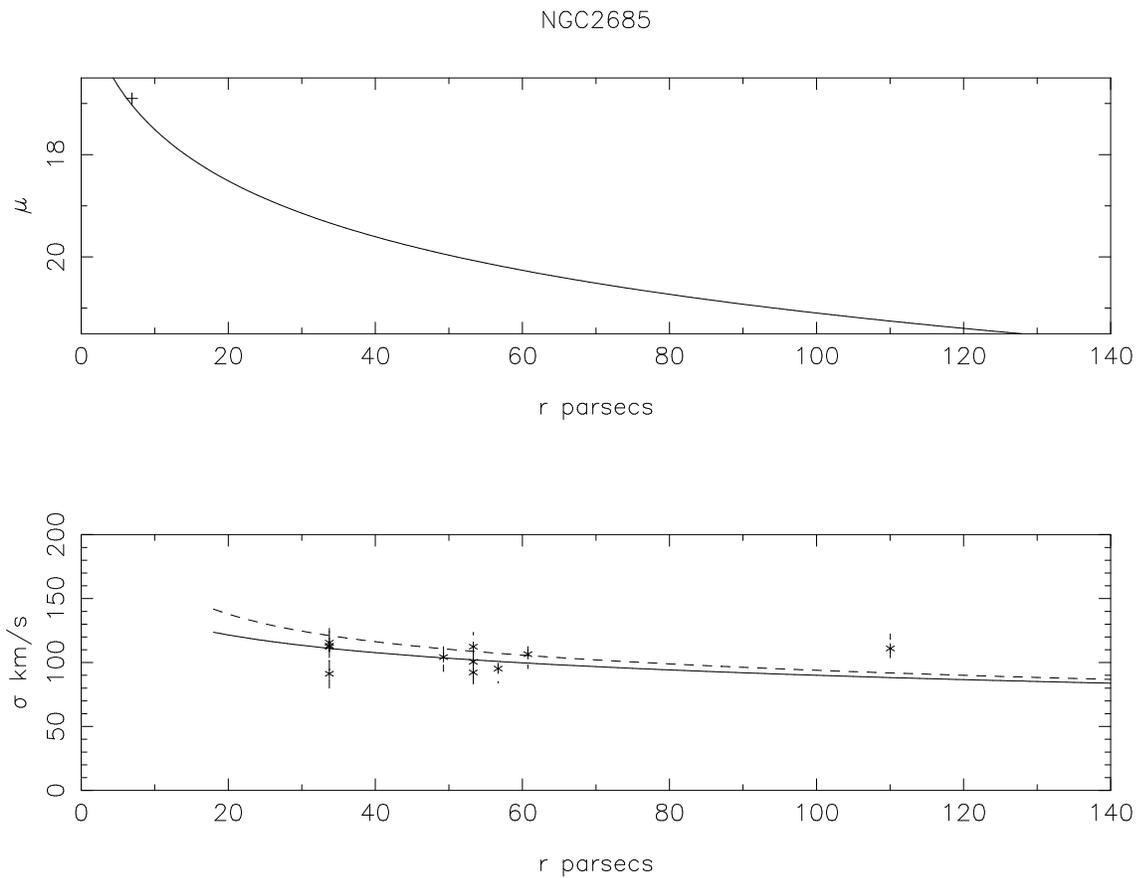}   
\caption{ $top$ NGC 2685's nuker profile; $bottom$ a fit to the TripleSpec
data without and with (dashed line) a 1.2 $\times$ 10$^7$ M$_\odot$ SMBH. }
\vspace*{3 cm}
\end{figure}

\begin{figure}[t]
\vspace*{-3 cm}
\includegraphics[scale=.2]{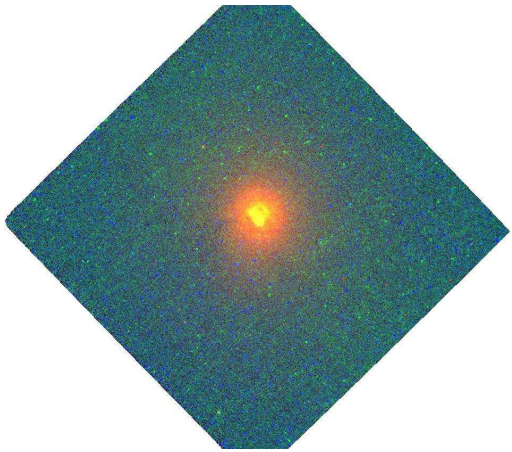}   
\caption{The HST WFPC2 PC image of the nucleus of NGC5273. The filters in RGB order are F547M/F300W/F218W.} 
\end{figure}

\begin{figure}
\includegraphics[width=.7\columnwidth,angle=-90]{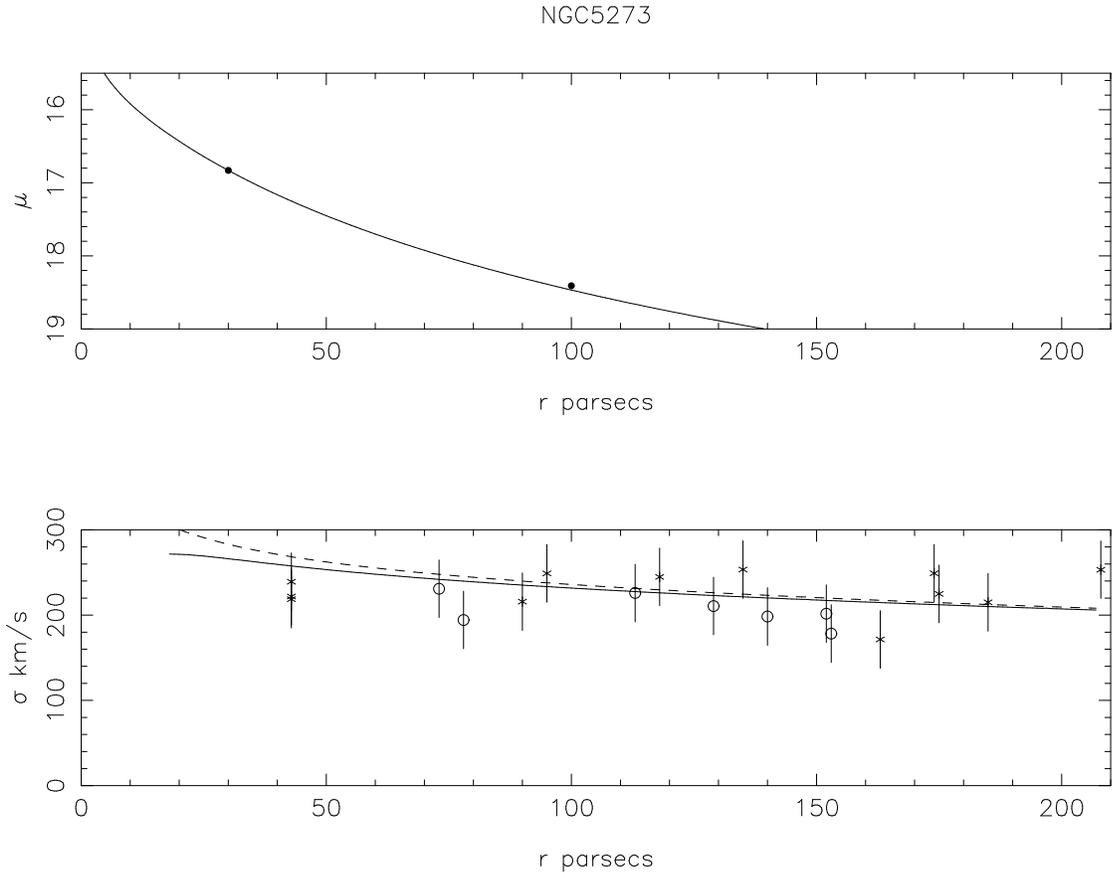}   
\caption{$top$ the two data points are from Mu\~noz Marin et al 2007; $bottom$
the dashed line has a SMBH of 1 $\times$ 10$^8$ M$_\odot$. Error bars are proportional to the cross correlation peak heights in Table 1.}
\vspace*{3 cm}
\end{figure}


\begin{figure}[t]
\vspace*{-3 cm}
\includegraphics[scale=.25]{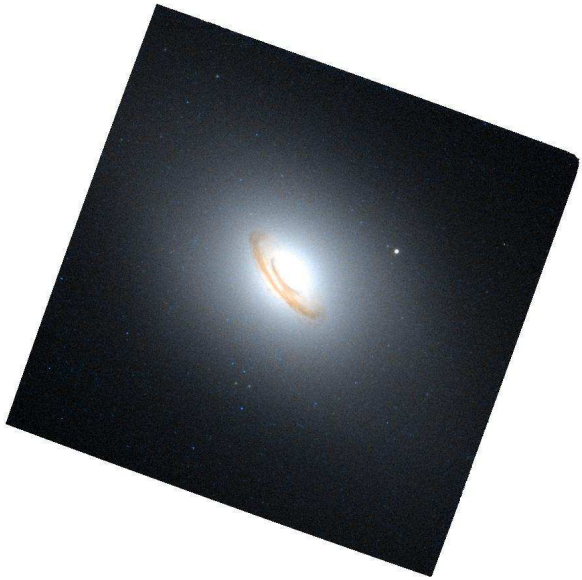}   
\caption{The HST WFPC2 PC image of the nucleus of NGC5838. The blue and red filters
are F450W and F814W. NASA: Hubble Legacy Archive.}
\end{figure}
\begin{figure}
\includegraphics[width=.7\columnwidth,angle=-90]{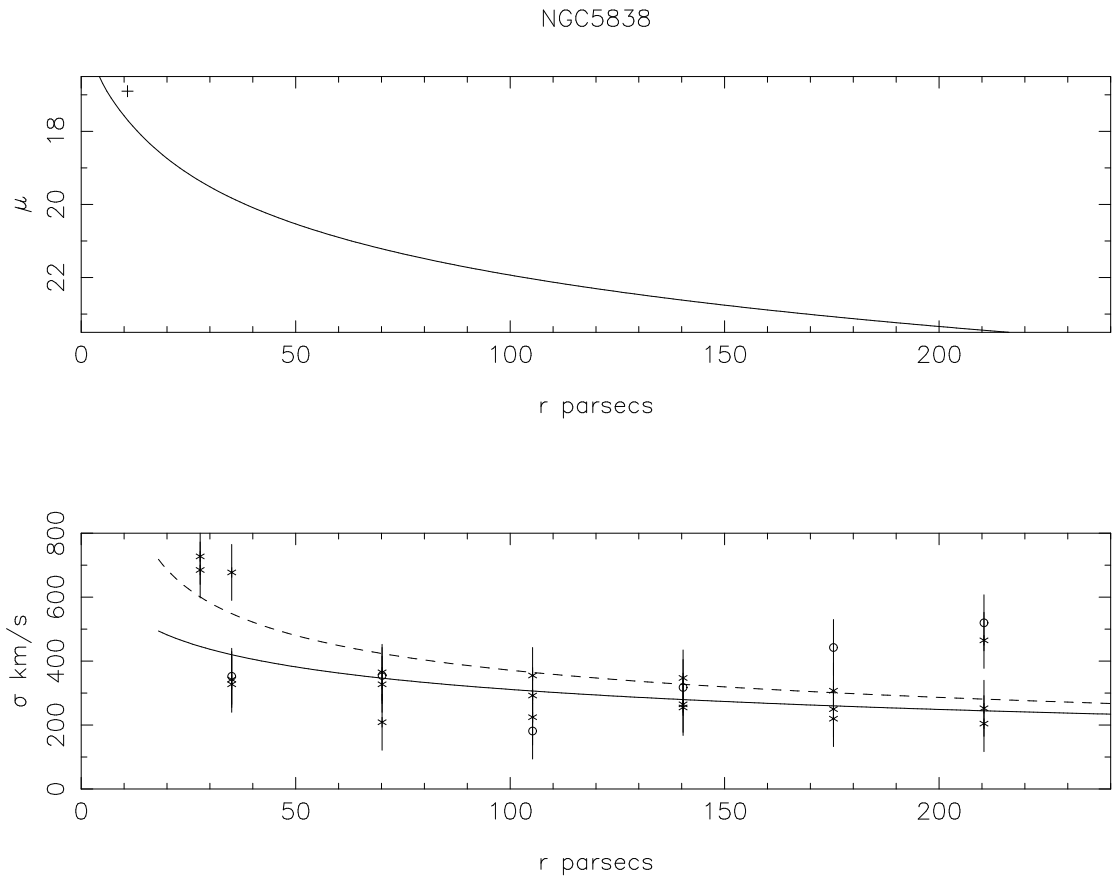}   
\caption{ The lower figure compares a fit to NGC 5838 with M/L = 30 (solid line) and adding a SMBH with mass 1 $\times$ 10$^8$ M$_\odot$ (dashed line).}
\end{figure}

\begin{figure}
\moveright 15mm \hbox{\includegraphics[width=.7\columnwidth,angle=-90]{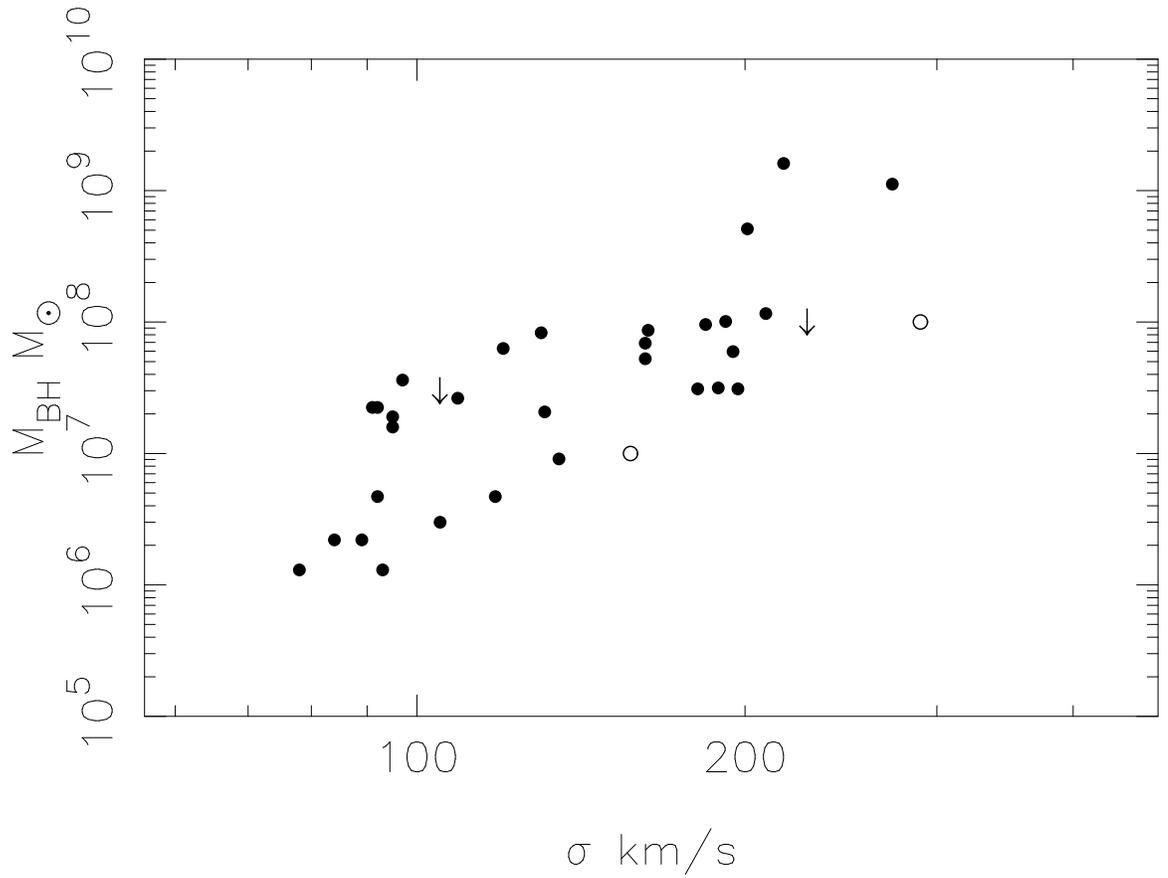}}
\caption{ Our two detections and two upper limits in the Magorrian diagram
of Bentz et al (2014).}
\end{figure}

\begin{figure}
\moveright 15mm \hbox{\includegraphics[width=.7\columnwidth,angle=-90]{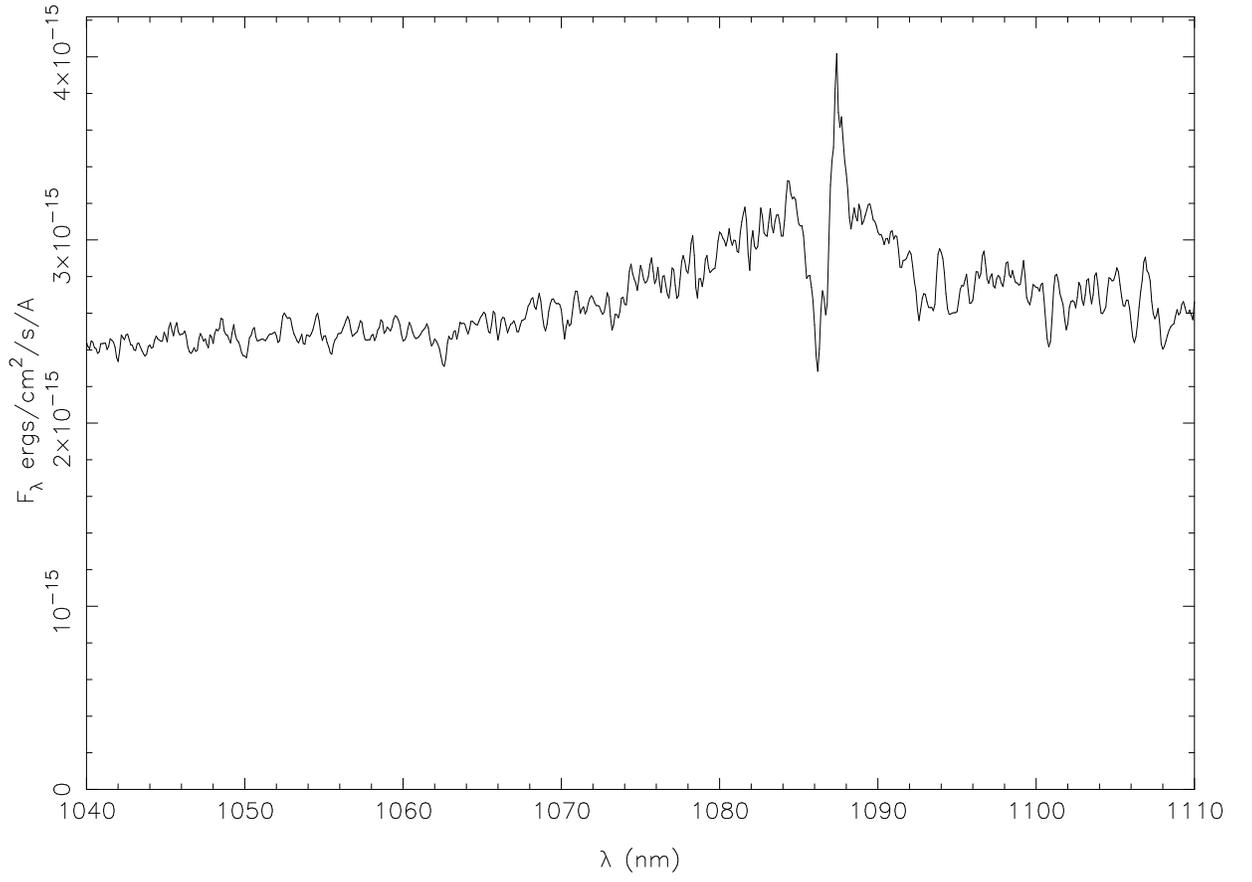}}
\caption{The He I 10830\AA~ region of NGC 5273.}
\end{figure}

\end{document}